\documentclass[journal]{IEEEtran}
\usepackage{amsmath,amssymb}
\usepackage{graphicx}
\usepackage[caption=false,font=footnotesize]{subfig}
\usepackage{cite}
\graphicspath{{./}}

\begin{document}
\title{Two-Dimensional Magnetometer Calibration Using Inertial Sensors for Heading Estimation}
\author{Mohammad Sadeghi, Rouhollah Amiri
	\thanks{Mohamamd Sadeghi is with School of Advanced Technologies, Iran University of Science and Technology, Tehran, Iran (e-mail: sadeghy@iust.ac.ir). Rouhollah Amiri is with Department of Electrical Engineering, Sharif University of Technology, Tehran, Iran, (e-mail: amiri@sharif.edu.)}}
\maketitle

\begin{abstract}
Magnetometers have numerous applications, one of the most important being heading determination. In practice, magnetometer measurements are affected by sensor manufacturing errors and by hard-iron and soft-iron effects, and therefore calibration is required for reliable heading estimation. Many conventional three-dimensional calibration methods require sufficiently rich multi-axis rotational excitation, which is not feasible in some applications, particularly for in-situ or online calibration. This paper proposes a two-dimensional magnetometer calibration method aided by inertial sensors. The proposed method is simple, closed-form, and non-iterative, and uses a single-axis rotational motion together with measured roll and pitch angles. After calibration, the corrected magnetic measurements are used for heading determination. Simulation and experimental results demonstrate the effectiveness of the proposed method under limited-rotation conditions.
\end{abstract}


\section{Introduction}
Accurate attitude and heading determination is essential in a wide range of applications, including unmanned aerial vehicles (UAVs), marine vehicles, and satellites \cite{ref1,ref2,ref3,ref4}. The attitude of a vehicle is commonly represented by three angles, namely roll, pitch, and yaw. The roll and pitch angles can be measured using different sensors, including accelerometers. Once the yaw angle, or equivalently the heading, is determined, the complete attitude of the vehicle can be obtained.

Several approaches can be used for heading determination, including magnetic compasses, GNSS, gyroscopic compasses, and celestial objects or star trackers \cite{ref5}. A gyroscopic compass is an accurate method; however, it is relatively time-consuming and costly. Celestial navigation or star trackers also involve considerable complexity and can only be used during suitable periods of the day and under appropriate operating conditions. GNSS is an effective method with acceptable accuracy and cost; however, in many situations, its signals are vulnerable to interference and jamming, and therefore it cannot always be used. In environments where GNSS signals are disrupted, magnetometers provide a suitable alternative. Even when GNSS is available, magnetometers can be employed as an independent sensor in a multi-sensor data-fusion framework. Therefore, owing to their small size and weight, low power consumption, and low cost, magnetometers are widely used either as standalone sensors or as auxiliary sensors in navigation and attitude-determination systems.

Magnetometer-based heading determination provides a simple and low-cost solution and can achieve acceptable accuracy. In the absence of disturbances and sensor imperfections, a magnetometer measures a fixed local geomagnetic-field vector, from which magnetic north can be determined. However, the north direction commonly used in attitude determination and navigation is geographic north. The angular difference between magnetic north and geographic north, known as magnetic declination, depends on geographic location. Geomagnetic reference models, such as the International Geomagnetic Reference Field (IGRF), can be used to account for this difference and obtain geographic north from magnetic heading.

In practical applications, however, magnetometer measurements are affected by several error sources. Some of these errors are related to external environmental factors, such as hard-iron and soft-iron effects, whereas others are related to the sensor hardware and manufacturing process, such as axis nonorthogonality, scale-factor errors, and misalignment. In magnetometer calibration models, these effects are commonly represented by a bias vector and a distortion matrix acting on the ideal magnetic-field vector \cite{ref6,ref7,ref8}. Magnetometer-based heading determination therefore requires compensation for these errors.

From a broad perspective, existing magnetometer calibration methods can be divided into two main categories: attitude-independent methods \cite{ref6,ref7,ref8,ref9} and methods based on inertial sensors \cite{ref10,ref11,ref12,ref13}. Attitude-independent methods exploit the invariance of the magnitude of the magnetic field. That is, measurements collected at different sensor orientations should ideally lie on a sphere with a radius corresponding to the magnitude of the local magnetic field. Therefore, these methods mainly use the magnitude of the magnetic-field vector rather than its direction. In methods based on inertial sensors, the variation of the reference magnetic-field vector can be compared with the measured magnetic vector. Therefore, these methods make use of the magnetic-field vector, including both its magnitude and direction.

With these considerations, attitude-independent methods for estimating ellipsoidal calibration parameters generally require sufficiently wide and diverse rotations about all three axes. If the measurements are collected only within a single plane or over a limited angular range, the coefficient matrix can become ill-conditioned, resulting in an inaccurate estimation of the calibration parameters \cite{ref6,ref7,ref8,ref9}. In contrast, methods based on inertial sensors can perform calibration even when the available rotations are limited. When sufficiently complete three-dimensional rotations are available, however, higher calibration accuracy can be obtained compared with attitude-independent methods.

Many magnetometer calibration methods require a laboratory environment and accurate calibration equipment, which limits their use in practical operating conditions. Magnetometer calibration compensates for both external magnetic disturbances and sensor-related errors. If the external environment changes, different hard-iron and soft-iron effects may be imposed on the sensor, and a calibration obtained in the laboratory may no longer remain valid \cite{ref14,ref15}. Therefore, for some practical applications, in-situ or online calibration is necessary.

For example, C\c{e}tin \cite{ref16} developed a recursive, attitude-independent calibration framework specifically for spinning aerospace vehicles. In \cite{ref17}, an in-field calibration method was presented for an autonomous underwater vehicle by reducing the number of calibration parameters from 12 to 9. Canciani demonstrated the importance of online calibration for magnetic navigation of an operational F-16 aircraft. The magnetic environment of the aircraft contained disturbances several times larger than those observed in experimental flight platforms, and a combination of manual and online calibration was used to achieve accurate navigation performance \cite{ref18}. Therefore, magnetometer calibration is not necessarily a one-time laboratory operation; for some aerospace and navigation systems, calibration must be applicable during normal vehicle operation and under changing magnetic conditions.

From a practical viewpoint, magnetometer calibration is often classified as either three-dimensional or two-dimensional calibration. In two-dimensional calibration, the sensor is rotated in a plane about a single axis, whereas in three-dimensional calibration, the sensor is rotated through a three-dimensional space. The measurements obtained during two-dimensional calibration approximately form an ellipse, while those obtained during three-dimensional calibration form an ellipsoid.

In many practical applications, a three-dimensional online calibration procedure is not feasible, and only two-dimensional calibration can be employed. This issue was investigated in \cite{ref9}, where a two-dimensional Analytical Extended Two-Step (2D-AETS) calibration method was introduced for applications in which only one rotational degree of freedom is available. The calibration is obtained in closed form rather than through an iterative optimization procedure. This method provides a relatively good approximation for pitch and roll angles smaller than approximately $\pm 10^\circ$. However, because attitude-independent methods rely on the formation of a sphere, the required spherical measurement distribution is not formed under such two-dimensional conditions, and consequently the performance of the method can deviate from the expected performance.

In other words, a purely planar calibration model implicitly assumes that the sensor remains at a constant tilt during the calibration rotation. Satisfying this assumption in practical operating environments can be difficult. Variations in roll and pitch change the projection of the geomagnetic-field vector onto the sensor axes and can therefore become coupled with the calibration parameters. Ignoring these effects may lead to biased calibration estimates, even when the intended calibration motion is nominally two-dimensional. The use of inertial measurements provides a means of obtaining this additional attitude information. Previous studies have demonstrated the advantages of combining magnetometer measurements with inertial-sensor information for calibration and subsequent attitude estimation \cite{ref10}.

In \cite{ref19}, the problem of two-dimensional magnetometer calibration using inertial sensors was investigated. However, it was assumed that the nonorthogonality, scale-factor, and bias errors had already been calibrated, and the objective was only to calibrate the misalignment error, i.e., to remove the misalignment angle between the three-axis magnetometer coordinate system and the accelerometer/INS coordinate system. In addition to requiring factory calibration before applying the calibration procedure, the assumptions of \cite{ref19} may not hold in environments with hard-iron effects, which limits its applicability.

In \cite{ref20}, a two-dimensional magnetometer calibration method based on roll and pitch measurements was proposed. In this method, it is assumed that, in addition to the total magnetic-field magnitude, the horizontal and vertical components of the magnetic field remain constant. A cost function is then formulated, and the calibration problem is solved using the Levenberg--Marquardt optimization method with an initial estimate. The paper also assumes an upper-triangular distortion matrix, which may restrict the method's applicability to more general calibration configurations. Furthermore, the iterative optimization algorithm is more complex and time-consuming and may not be suitable for some real-time applications. More importantly, the algorithm depends on the initial estimate. Since the problem is solved iteratively, an inappropriate initial estimate may prevent convergence to the desired solution.

In this paper, we introduce a simple and effective closed-form two-dimensional magnetometer calibration method that combines single-axis rotational excitation with measured roll and pitch information. The proposed method does not require an initial parameter estimate or an iterative optimization algorithm, which makes the calibration procedure simple and non-iterative while retaining the practical advantages of two-dimensional calibration for in-situ and online applications. After calibration, the corrected magnetic measurements are used to estimate the heading.

It should be emphasized that the proposed method does not perform a full three-dimensional magnetometer calibration. Instead, it estimates the calibration parameters required for two-dimensional heading determination using measurements acquired during a single-axis rotational motion.

The remainder of this paper is organized as follows. Section II presents the magnetometer measurement model. Section III develops the proposed two-dimensional calibration method using single-axis rotational excitation and roll and pitch measurements, followed by the heading-estimation procedure. Section IV evaluates the proposed method through numerical simulations, including symmetric and asymmetric distortion-matrix cases and comparison with the 2D-AETS method. Section V presents the experimental setup, calibration procedure, and heading-estimation results for two test scenarios. Finally, Section VI concludes the paper, and Appendix A describes the ellipse-fitting procedure used in the calibration algorithm.

\section{System Model}
The navigation frame is defined as a North-East-Down (NED) frame, in which the $x$-, $y$-, and $z$-axes point toward North, East, and Down, respectively. The body frame is defined as a Forward-Right-Down (FRD) frame, where the $x_b$-, $y_b$-, and $z_b$-axes point forward, right, and down, respectively. The body and navigation frames are assumed to be aligned at the reference orientation.

The measurements obtained from a magnetometer are
affected by several error sources. Some of these errors originate
from the sensor manufacturing process, including bias, scale-factor, and misalignment errors \cite{ref21}. Other errors are associated
with environmental effects, which are commonly modeled as
hard-iron and soft-iron effects \cite{ref22}. Bias and scale-factor errors are
characteristics of an individual sensor, whereas misalignment
errors arise from manufacturing tolerances in the sensor
assembly. In contrast, hard-iron and soft-iron effects depend
on the surrounding environment and may vary under different
operating conditions.

A sensor bias produces a constant shift in the measured
output and can be modeled by a three-dimensional bias vector. 
The manufacturer attempts to make the gains of the
three magnetometer axes equal through factory calibration;
however, residual errors generally remain. The scale-factor
error represents the ratio between the actual magnetic-field
component and the corresponding measured component along
each axis and can be modeled by a diagonal matrix.
Misalignment and nonorthogonality between the sensor axes,
which originate from manufacturing tolerances, can be represented by a matrix.
The hard-iron effect is caused by hard-magnetic materials
with a wide hysteresis loop, such as permanent magnets, motors,
and hardened steel, which introduce an approximately constant
magnetic field in the sensor environment. The hard-iron effect
shifts the center of the measured ellipse and can therefore be
modeled by a bias vector. The soft-iron effect is caused by
soft ferromagnetic materials with a narrow hysteresis loop, such
as iron frames and cables carrying alternating current. These
materials modify the magnetic field and distort the measured
ellipse.

The general magnetometer measurement model is given by \cite{ref22}
\begin{equation}
\mathbf{m}=W_e R_x(\phi)R_y(\theta)R_z(\psi)b_0
\begin{bmatrix}\cos\delta\\0\\\sin\delta\end{bmatrix}+\mathbf{v}+\mathbf{n},
\label{eq:model_we}
\end{equation}
where $b_0$ denotes the magnitude of the magnetic field at the operating location, $\delta$ denotes the magnetic inclination, which is assumed to be known from a geomagnetic reference model, $W_e$ is a $3\times3$ distortion matrix representing the
combined effects of soft iron, axis nonorthogonality, misalignment, and scale factor, $\mathbf{v}$ is the combined bias and hard-iron vector, and $\mathbf{n}$ is a zero-mean Gaussian measurement-noise vector. Since $b_0$ is constant for the considered calibration environment, its effect is absorbed into the distortion matrix by defining
\begin{equation}
W=b_0W_e.
\label{eq:Wabsorb}
\end{equation}
The matrix $W$ is a field-scaled distortion matrix that incorporates both the local magnetic-field magnitude and the linear sensor/environmental distortion.
Therefore, the measurement model used in the proposed calibration formulation becomes
\begin{equation}
\mathbf{m}=W R_x(\phi)R_y(\theta)R_z(\psi)
\begin{bmatrix}\cos\delta\\0\\\sin\delta\end{bmatrix}+\mathbf{v}+\mathbf{n}.
\label{eq:model}
\end{equation}
The bias vector is
\begin{equation}
\mathbf{v}=\mathbf{v}^{\mathrm{bias}}+\mathbf{v}^{\mathrm{hard}}=\begin{bmatrix}v_x&v_y&v_z\end{bmatrix}^{T}.
\end{equation}

Various models for the matrix $W$ have been proposed in the literature. Some studies assume a 3-parameter model with a diagonal matrix \cite{ref15}, others a 6-parameter model with a symmetric matrix \cite{ref8}, some a 6-parameter model with a lower or upper triangular matrix \cite{ref6,ref9}, and others a general 9-parameter model \cite{ref10,ref23}. In this paper, we adopt the 6-parameter symmetric matrix model.

Thus, the distortion matrix used in this paper is represented as
\begin{equation}
W=\begin{bmatrix}a&d&e\\d&b&f\\e&f&c\end{bmatrix}.
\end{equation}
In general, $W$ may be asymmetric in practical
systems. The effect of an asymmetric
$W$ is nevertheless investigated in the simulation section to
evaluate the robustness of the proposed method. 

The rotation matrices are
\begin{equation}
R_x(\phi)=\begin{bmatrix}1&0&0\\0&\cos\phi&\sin\phi\\0&-\sin\phi&\cos\phi\end{bmatrix},
\end{equation}
\begin{equation}
R_y(\theta)=\begin{bmatrix}\cos\theta&0&-\sin\theta\\0&1&0\\\sin\theta&0&\cos\theta\end{bmatrix},
\end{equation}
\begin{equation}
R_z(\psi)=\begin{bmatrix}\cos\psi&\sin\psi&0\\-\sin\psi&\cos\psi&0\\0&0&1\end{bmatrix}.
\end{equation}
Assuming $W$, $\mathbf{v}$, and the pitch and roll angles are known, the calibrated vector is
\begin{equation}
\mathbf{m}_f=R_y(-\theta)R_x(-\phi)W^{-1}(\mathbf{m}-\mathbf{v})=R_z(\psi)\begin{bmatrix}\cos\delta\\0\\\sin\delta\end{bmatrix}.
\end{equation}
where $\mathbf{m}_f$ denotes the calibrated and tilt-compensated magnetic-field vector.
Thus,
\begin{equation}
m_f^x=\cos\psi\cos\delta,\qquad m_f^y=-\sin\psi\cos\delta,
\end{equation}
The yaw angle is therefore obtained from
\begin{equation}
\psi=\operatorname{atan2}(-m_f^y,m_f^x).
\end{equation}

\section{Two-Dimensional Magnetometer Calibration}
The magnetometer is rotated through one complete $360^\circ$ rotation about the $z$-axis. During the calibration stage, the pitch and roll angles, $\theta$ and $\phi$, are measured using an accelerometer installed next to the magnetometer. The corresponding axes of the accelerometer and magnetometer are assumed to be mechanically aligned. The accelerometer is operated under a quasi-static condition so that roll and pitch can be obtained from the gravity vector. The proposed formulation uses the minimum two magnetic measurement channels required for two-dimensional heading calibration, namely $m_x$ and $m_y$. The $m_z$ channel is therefore not included in the proposed calibration formulation. The calibration is performed approximately on a horizontal plane, so the roll and pitch angles remain relatively small during calibration. Larger roll and pitch angles may, however, be present during the subsequent test stage.

It should be emphasized that the proposed method does not perform a full three-dimensional magnetometer calibration. Instead, it estimates the calibration parameters required for two-dimensional heading determination using measurements acquired during a single-axis rotational motion.

\subsection{First-Stage Estimation}
Accordingly, the $x$-axis measurement is expressed as follows:
\begin{equation}
\begin{aligned}
m_x={}&a\cos\theta\cos\psi\cos\delta-a\sin\theta\sin\delta\\
&-d\cos\psi\sin\theta\sin\phi\cos\delta-d\cos\phi\sin\psi\cos\delta\\
&+d\cos\theta\sin\phi\sin\delta+e\sin\theta\cos\phi\cos\psi\cos\delta\\
&+e\sin\phi\sin\psi\cos\delta+e\cos\theta\cos\phi\sin\delta+v_x.
\end{aligned}
\end{equation}
Given that the angles $\theta$ and $\phi$ are near zero, the approximations $\cos\theta\simeq1$, $\cos\phi\simeq1$, and $\sin\theta\sin\phi\simeq0$ can be considered. Furthermore, since $\sin\theta\ll\cos\theta$ and $\sin\phi\ll\cos\phi$, the term $e\sin\theta\cos\phi\cos\psi\cos\delta$ can be neglected compared with $a\cos\theta\cos\psi\cos\delta$, and the term $e\sin\phi\sin\psi\cos\delta$ can be neglected compared with $d\cos\phi\sin\psi\cos\delta$. Using these approximations, the expression for $m_x$ can be written as
\begin{equation}
m_x\simeq a\cos\delta\cos\psi-d\cos\delta\sin\psi+v_x',
\end{equation}
where
\begin{equation}
v_x'=-a\sin\theta\sin\delta+d\cos\theta\sin\phi\sin\delta+e\sin\delta+v_x.
\end{equation}
The terms independent of $\theta$ and $\phi$ are considered as a new bias term,
\begin{equation}
v_x^n=e\sin\delta+v_x.
\end{equation}
Similarly,
\begin{equation}
v_y'=-d\sin\theta\sin\delta+b\cos\theta\sin\phi\sin\delta+v_y^n,
\end{equation}
where
\begin{equation}
v_y^n=f\sin\delta+v_y.
\end{equation}
Therefore,
\begin{equation}
m_x\simeq(a\cos\delta)\cos\psi+(-d\cos\delta)\sin\psi+v_x',
\end{equation}
\begin{equation}
m_y\simeq(d\cos\delta)\cos\psi+(-b\cos\delta)\sin\psi+v_y'.
\end{equation}

The measured $m_x$ and $m_y$ data form an ellipse. Assuming that $\cos\delta$ is known, fitting an ellipse to the measurement set provides estimates of $a$, $b$, $d$, $v_x'$, and $v_y'$. The ellipse-fitting procedure is described in Appendix A. The five geometric fitting parameters are $a_e$, $b_e$, $\phi_e$, $x_{0e}$, and $y_{0e}$. The bias estimates are
\begin{equation}
\widehat v_x'=x_{0e},\qquad \widehat v_y'=y_{0e}.
\end{equation}
The parameters $a$, $b$, and $d$ can now be estimated as follows:
\begin{equation}
W_h=\begin{bmatrix}\widehat a&\widehat d\\\widehat d&\widehat b\end{bmatrix}=\frac{R_e\Lambda_eR_e^T}{\cos\delta},
\end{equation}
where $(\cdot)^T$ denotes the transpose operator and
\begin{equation}
\Lambda_e=\begin{bmatrix}a_e&0\\0&b_e\end{bmatrix},\qquad
R_e=\begin{bmatrix}\cos\phi_e&\sin\phi_e\\-\sin\phi_e&\cos\phi_e\end{bmatrix}.
\end{equation}

It should be noted that the proposed method is not applicable at the magnetic poles, where the magnetic inclination approaches $+90^\circ$ or $-90^\circ$. In this condition, $\cos\delta$ approaches zero and the horizontal component of the geomagnetic field becomes very small, which prevents reliable estimation of the horizontal calibration parameters.

The remaining bias terms are estimated as
\begin{equation}
\widehat v_x^n=\widehat v_x'+\widehat a\sin\bar\theta\sin\delta-\widehat d\cos\bar\theta\sin\bar\phi\sin\delta,
\end{equation}
\begin{equation}
\widehat v_y^n=\widehat v_y'+\widehat d\sin\bar\theta\sin\delta-\widehat b\cos\bar\theta\sin\bar\phi\sin\delta,
\end{equation}
where
\begin{equation}
\bar\theta=\frac{1}{N}\sum_{i=1}^{N}\theta_i,\qquad \bar\phi=\frac{1}{N}\sum_{i=1}^{N}\phi_i.
\end{equation}

\subsection{Second-Stage Estimation}
In the first stage, the small-angle approximation and approximately constant $v_x'$ and $v_y'$ were used. In the second stage, the measured roll and pitch variations are incorporated more directly by defining
\begin{equation}
m_1=h_{11}\cos\psi+h_{12}\sin\psi\simeq m_x-v_x',
\end{equation}
\begin{equation}
m_2=h_{21}\cos\psi+h_{22}\sin\psi\simeq m_y-v_y',
\end{equation}
where
\begin{equation}
h_{11}=\widehat a\cos\theta\cos\delta,\quad h_{12}=-\widehat d\cos\phi\cos\delta,
\end{equation}
\begin{equation}
h_{21}=\widehat d\cos\theta\cos\delta,\quad h_{22}=-\widehat b\cos\phi\cos\delta.
\end{equation}
Thus,
\begin{equation}
\begin{bmatrix}h_{11}&h_{12}\\h_{21}&h_{22}\end{bmatrix}\begin{bmatrix}\cos\psi\\\sin\psi\end{bmatrix}=\begin{bmatrix}m_1\\m_2\end{bmatrix},
\end{equation}
so that
\begin{equation}
\begin{bmatrix}\cos\psi\\\sin\psi\end{bmatrix}=\begin{bmatrix}h_{11}&h_{12}\\h_{21}&h_{22}\end{bmatrix}^{-1}\begin{bmatrix}m_1\\m_2\end{bmatrix}.
\end{equation}
The estimates of $m_1$ and $m_2$ are
\begin{equation}
\widehat m_1=m_x-\widehat v_x^n+\widehat a\sin\theta\sin\delta-\widehat d\cos\theta\sin\phi\sin\delta,
\end{equation}
\begin{equation}
\widehat m_2=m_y-\widehat v_y^n+\widehat d\sin\theta\sin\delta-\widehat b\cos\theta\sin\phi\sin\delta.
\end{equation}
The estimates $T_1$ and $T_2$ of $\cos\psi$ and $\sin\psi$ are
\begin{equation}
\begin{bmatrix}T_1\\T_2\end{bmatrix}=\frac{1}{h_{11}h_{22}-h_{12}h_{21}}\begin{bmatrix}h_{22}&-h_{12}\\-h_{21}&h_{11}\end{bmatrix}\begin{bmatrix}\widehat m_1\\\widehat m_2\end{bmatrix}.
\end{equation}
If calibration is complete, the $(T_1,T_2)$ data form a circle. The residual correction matrix is obtained from a second ellipse fit as
\begin{equation}
W_{h2}=R_{e2}\begin{bmatrix}a_{e2}&0\\0&b_{e2}\end{bmatrix}R_{e2}^{T},
\end{equation}
where $R_{e2}$ is the rotation matrix associated with the fitted ellipse. The final horizontal matrix is directly constructed as
\begin{equation}
W_h^F=W_hW_{h2}=\begin{bmatrix}a^F&d_{12}^F\\d_{21}^F&b^F\end{bmatrix},
\end{equation}
with
\begin{equation}
d^F=\frac{d_{21}^F+d_{12}^F}{2}.
\end{equation}

To assess the accuracy of the calibration, the two estimated vectors $T_1$ and $T_2$ can be updated as follows:
\begin{equation}
\begin{bmatrix}
T_1^F\\
T_2^F
\end{bmatrix}
=\frac{1}{h_{11}^Fh_{22}^F-h_{12}^Fh_{21}^F}
\begin{bmatrix}
h_{22}^F&-h_{12}^F\\
-h_{21}^F&h_{11}^F
\end{bmatrix}
\begin{bmatrix}
\widehat m_1\\
\widehat m_2
\end{bmatrix}.
\end{equation}
where
\begin{equation}
h_{11}^F=a^F\cos\theta\cos\delta,\qquad h_{12}^F=-d_{12}^F\cos\phi\cos\delta,
\end{equation}
\begin{equation}
h_{21}^F=d_{21}^F\cos\theta\cos\delta,\qquad h_{22}^F=-b^F\cos\phi\cos\delta.
\end{equation}
We expect $T_1^F$ and $T_2^F$ to form a circle. A simple and suitable measure for evaluating the circularity of these vectors is defined as follows. This measure is referred to as the circle deviation:
\begin{equation}
\sigma_c=\sqrt{\frac{1}{N}\sum_{i=1}^{N}\left(T_1^F(i)^2+T_2^F(i)^2-1\right)^2}.
\end{equation}
A smaller value indicates better calibration, and the ideal value is zero.

\subsection{Heading Estimation}
After the calibration process, the parameters $v_x^n$, $v_y^n$, $a^F$, $b^F$, and $d^F$ are obtained. If the measurements $b_m^x$, $b_m^y$, $\theta_m$, and $\phi_m$ are available at any instant, the heading can be estimated using the following relation:
\begin{equation}
\begin{aligned}
\psi_i=\operatorname{atan2}\Big(& (m_1d^F-m_2a^F)\cos\theta_m,\\
& (m_1b^F-m_2d^F)\cos\phi_m\Big).
\end{aligned}
\end{equation}
where
\begin{equation}
m_1=b_m^x-v_x^n+a^F\sin\theta_m\sin\delta-d^F\cos\theta_m\sin\phi_m\sin\delta,
\end{equation}
\begin{equation}
m_2=b_m^y-v_y^n+d^F\sin\theta_m\sin\delta-b^F\cos\theta_m\sin\phi_m\sin\delta.
\end{equation}

It should be noted that, in all magnetometer-based heading-estimation methods, an approximate knowledge of the user's position, within a few tens of kilometers, is required. This is because the difference between magnetic north and geographic north depends on magnetic declination, which in turn depends on geographic location. Given an approximate user position, the magnetic inclination $\delta$ can also be obtained and treated as known with a certain accuracy. The effect of heading-estimation error caused by uncertainty in $\delta$ is investigated in the simulation.

\section{Simulation Results}
The proposed method is evaluated through numerical simulations. Two separate processes are considered: calibration and testing.

\subsection{Calibration Process}
During calibration, the magnetometer together with the inertial sensors is rotated through one complete $360^\circ$ rotation about the $z$-axis. Since the rotation plane is not exactly parallel to the horizontal plane, variations in roll and pitch occur. The roll and pitch variations are modeled using the following sinusoidal function:
\begin{equation}
\eta(t)=\alpha\sin(2\pi t+\gamma)+\beta,
\end{equation}
where $\eta$ denotes the roll or pitch angle. For roll, the parameters are $(\alpha,\beta,\gamma)=(1,-0.9,45)$, and for pitch they are $(\alpha,\beta,\gamma)=(2,0.3,17)$, with all angular quantities expressed in degrees. The angle measurement noise is $0.1^\circ$, the magnetometer measurement noise is $2$ mG, the magnetic inclination is $55^\circ$, and $N=1000$ samples are collected during a complete rotation.

\subsection{Test Scenario}
The roll angle of the magnetometer is assumed to be $-0.2^\circ$. The magnetometer undergoes one complete $360^\circ$ rotation, during which the pitch angle varies from $0^\circ$ to $8^\circ$. The angle-sensor noise is $0.1^\circ$, the magnetometer measurement noise is 2 mG, and 1000 samples are used.

\subsection{Case A: Symmetric Distortion Matrix}
\begin{figure}[!t]
\centering
\includegraphics[width=\columnwidth]{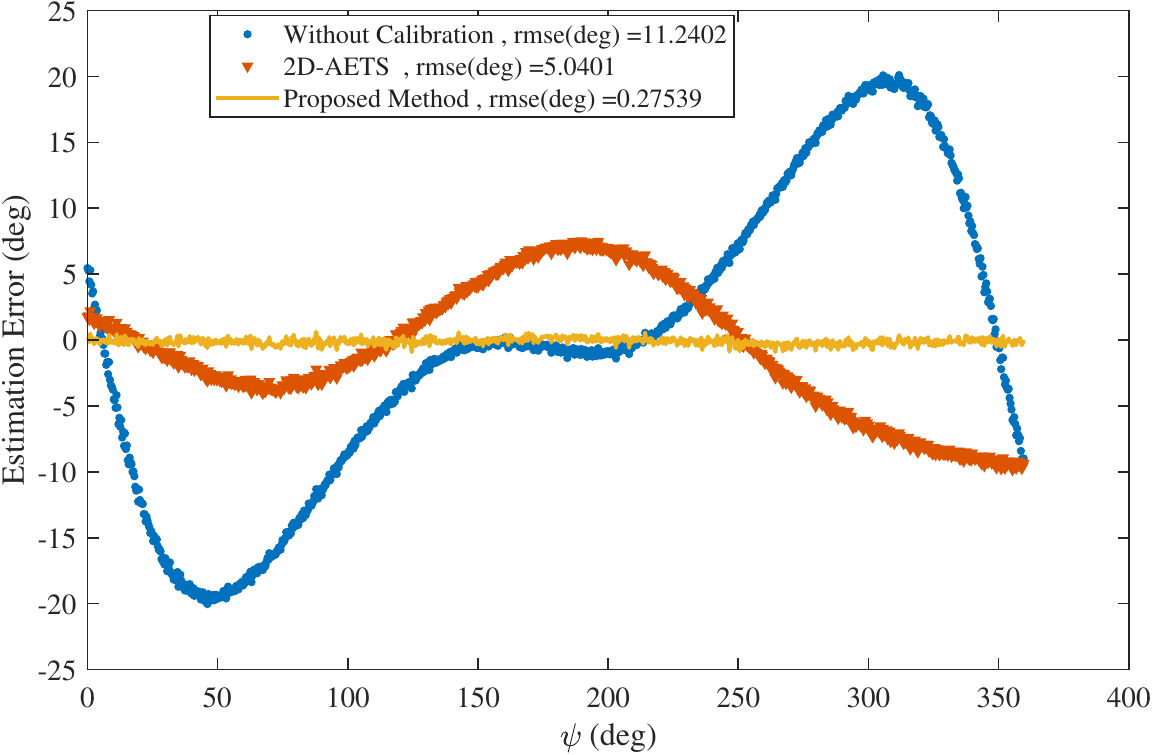}
\caption{Heading-estimation error for the symmetric-distortion-matrix simulation in Case A.}
\label{fig:sim_sym}
\end{figure}

The symmetric distortion matrix and bias vector are
\begin{equation}
W=\begin{bmatrix}1085&-30&20\\-30&804&-50\\20&-50&800\end{bmatrix},\qquad
\mathbf{v}=\begin{bmatrix}48&168&-400\end{bmatrix}^{T}.
\end{equation}
For the calibration data, the 2D-AETS method presented in \cite{ref9} and the proposed method are applied. The resulting heading estimates are compared with heading estimation without calibration. Fig.~\ref{fig:sim_sym} shows the heading-estimation error. Without calibration, the heading error is large. The 2D-AETS method partially compensates for the magnetometer disturbances, whereas the proposed method provides much better performance. The difference is partly attributed to the fact that 2D-AETS does not use inertial-sensor measurements during calibration. The obtained circle deviation is 0.0082.

\begin{figure}[!t]
\centering
\includegraphics[width=\columnwidth]{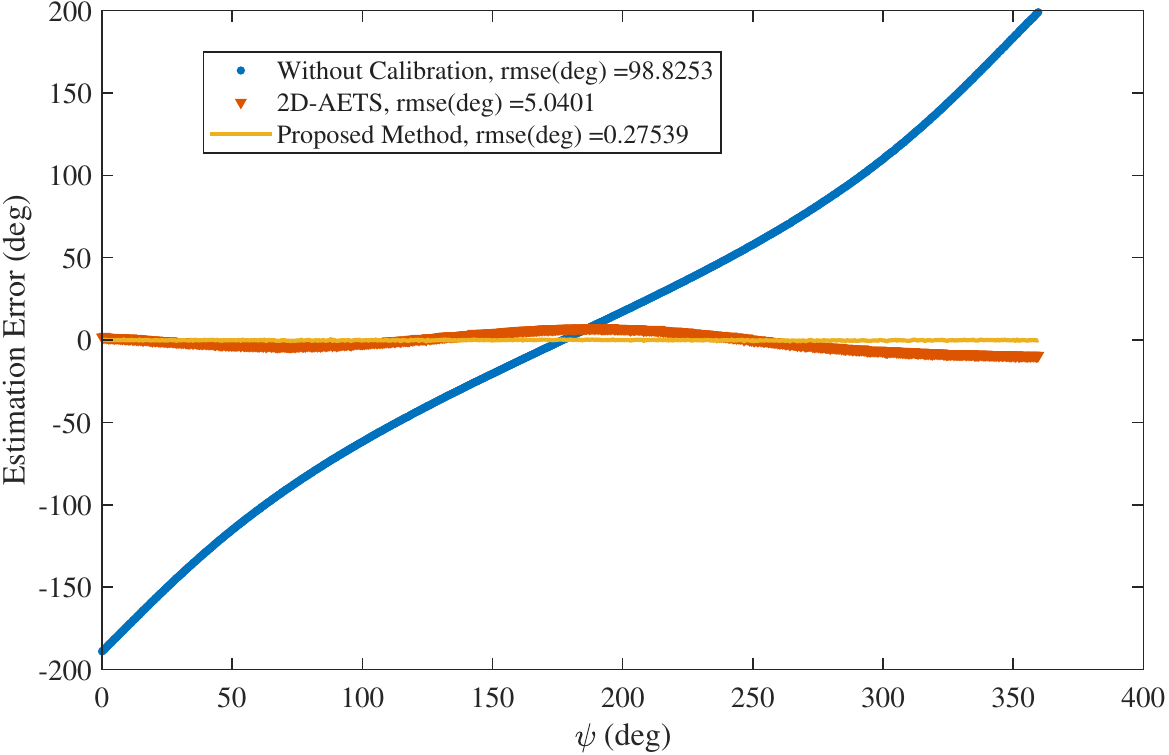}
\caption{Heading-estimation error with an increased bias vector representing a stronger hard-iron effect.}
\label{fig:sim_hard}
\end{figure}

The simulation is repeated with the same distortion matrix and a new bias vector,
\begin{equation}
\mathbf{v}=\begin{bmatrix}480&1680&-4000\end{bmatrix}^{T},
\end{equation}
to represent a more significant hard-iron effect. Fig.~\ref{fig:sim_hard} shows the results. In the presence of a significant hard-iron effect, heading estimation without calibration becomes ineffective. The 2D-AETS method partially compensates for the error; however, the proposed method continues to provide good performance.

\subsection{Effect of Uncertainty in Magnetic Inclination}
The effect of inaccurate knowledge of the magnetic inclination, $\delta$, on heading estimation is investigated in this section. As discussed previously, magnetometer-based heading estimation requires approximate knowledge of the sensor position. Based on a rule of thumb, if the sensor position is known with an accuracy better than 100 km, the magnetic declination can be obtained with an accuracy better than approximately $1^\circ$.

The simulation is performed using the parameters of Case A. The true magnetic inclination is considered to be $55^\circ$ and $30^\circ$, while the inclination value used in the heading-estimation test differs from the true value by $\Delta\delta$. The obtained heading-estimation errors are listed in Table~\ref{tab:delta_uncertainty}.

For $\delta=55^\circ$, the largest difference from the zero-inclination-error case occurs at $\Delta\delta=-2^\circ$, where the heading-estimation error increases from $0.27^\circ$ to $0.49^\circ$, corresponding to an additional error of $0.22^\circ$. For $\delta=30^\circ$, the corresponding values remain smaller. It should be noted that, in many practical applications, the sensor position can be available with an error smaller than 100 km; therefore, the heading-estimation error caused by inaccurate magnetic declination is expected to be limited.

\begin{table}[!t]
\centering
\caption{Heading-Estimation Error for Different Magnetic-Inclination Errors}
\label{tab:delta_uncertainty}
\scriptsize
\renewcommand{\arraystretch}{1.05}
\resizebox{\columnwidth}{!}{%
\begin{tabular}{c|rrrrrrrrr}
\hline
$\Delta\delta$ (deg) & $2$ & $1.5$ & $1$ & $0.5$ & $0$ & $-0.5$ & $-1$ & $-1.5$ & $-2$ \\
\hline
$\delta=55^\circ$ & 0.39 & 0.32 & 0.28 & 0.26 & 0.27 & 0.31 & 0.36 & 0.42 & 0.49 \\
$\delta=30^\circ$ & 0.22 & 0.21 & 0.21 & 0.21 & 0.21 & 0.22 & 0.24 & 0.26 & 0.29 \\
\hline
\end{tabular}%
}
\end{table}

\subsection{Case B: Asymmetric Distortion Matrix}
\begin{figure}[!t]
\centering
\includegraphics[width=\columnwidth]{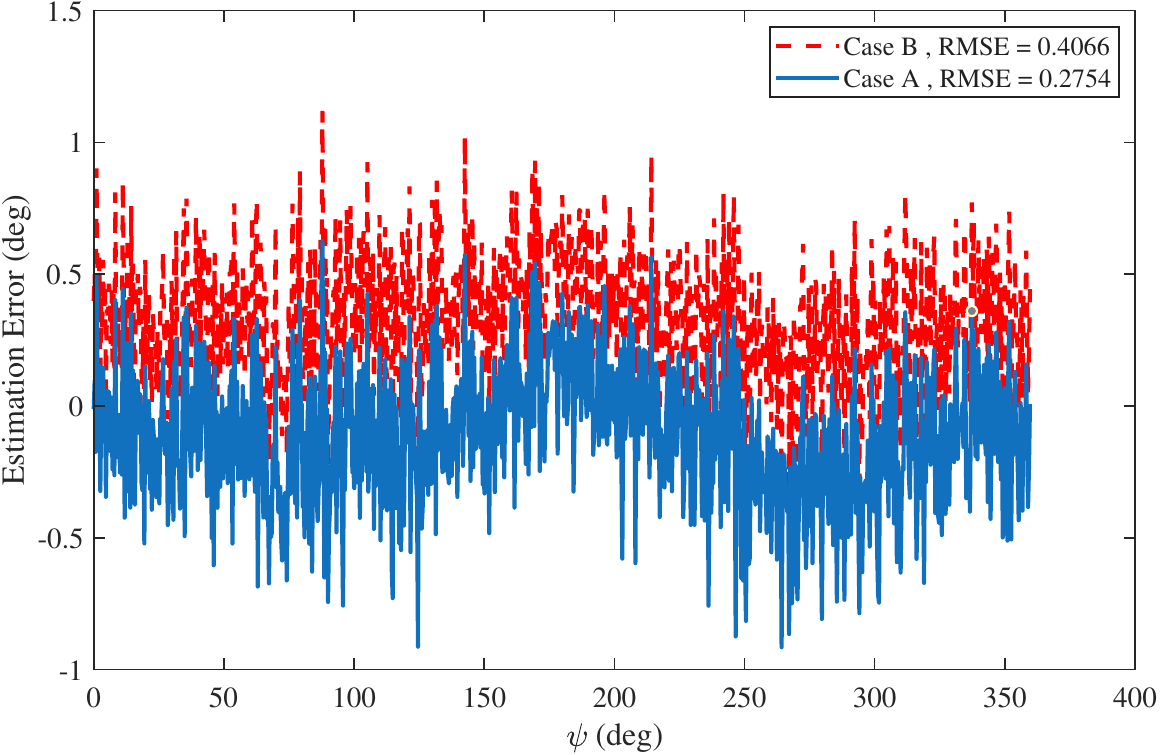}
\caption{Heading-estimation error for the Case A (symmetric-distortion-matrix) and Case B(asymmetric-distortion-matrix).}
\label{fig:sim_asym}
\end{figure}

The closed-form formulation is derived under the assumption that the distortion matrix is symmetric. To evaluate the performance under a more general condition, an asymmetric distortion matrix is considered:
\begin{equation}
\small
W=\begin{bmatrix}1085&-30&20\\-17&804&-23\\88&-50&800\end{bmatrix},\qquad
\mathbf{v}=\begin{bmatrix}480&1680&-4000\end{bmatrix}^{T}.
\normalsize
\end{equation}
The corresponding result is shown in Fig.~\ref{fig:sim_asym}. The proposed method continues to provide satisfactory performance with no significant degradation. A small bias is observed in the estimate, resulting in an RMSE of $0.4066^\circ$.


\section{Experimental Results}
The proposed method is further evaluated using experimental measurements. The measurement module consists of a QMC5883L magnetometer and an SCA100T accelerometer, which provide three-dimensional magnetic-field measurements together with roll and pitch angles.

\subsection{Calibration Procedure}
For the calibration, the sensors are mounted on a two-degree-of-freedom motion table. The table can rotate through $360^\circ$ in azimuth and through a limited range in elevation, both with the resolution of $0.05^\circ$. The rotation speed of moter is $0.5^\circ$/s.
The QMC5883L magnetometer and SCA100T inclinometer were sampled both at $1$ Hz. The magnetic inclination used by the algorithm was $\delta=55^\circ$, obtained from IGRF for the test location and date.

The mathematical equations developed in Section III are based on the assumption that the roll and pitch angles are small. Therefore, the calibration procedure should be performed around small roll and pitch angles. After calibration, however, the proposed method can be used during testing under non-horizontal operating conditions. For improved calibration accuracy, it is preferable to perform the $360^\circ$ rotation in both directions, i.e., one clockwise and one counterclockwise rotation. The measurement data should be acquired from the beginning of the motor rotation and stopped when the rotation is completed.

\begin{figure}[!t]
\centering
\subfloat[Before calibration and fitted ellipse.]{\includegraphics[width=0.47\columnwidth]{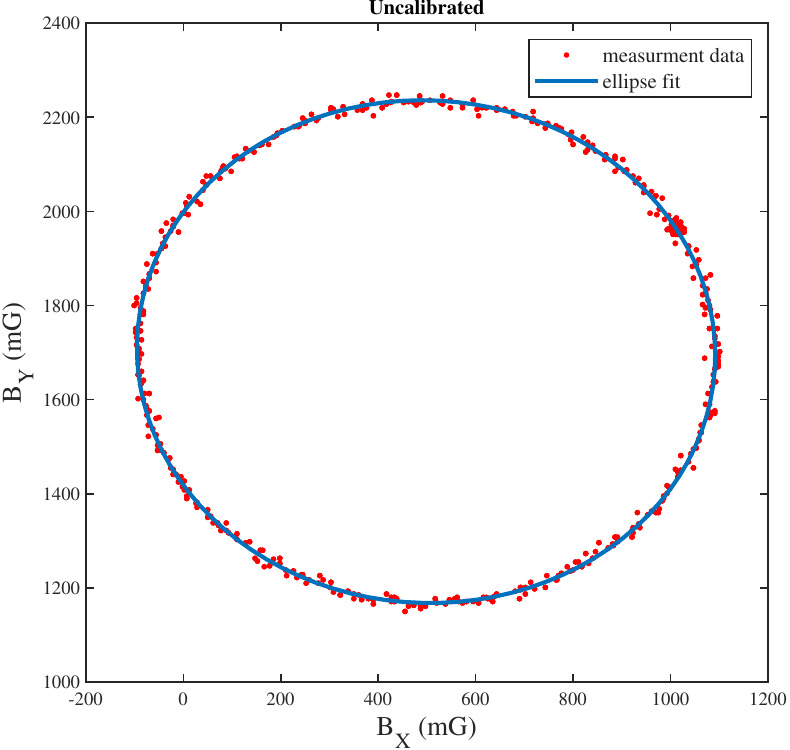}\label{fig:exp_cal_a}}
\hfill
\subfloat[After calibration and fitted unit circle.]{\includegraphics[width=0.47\columnwidth]{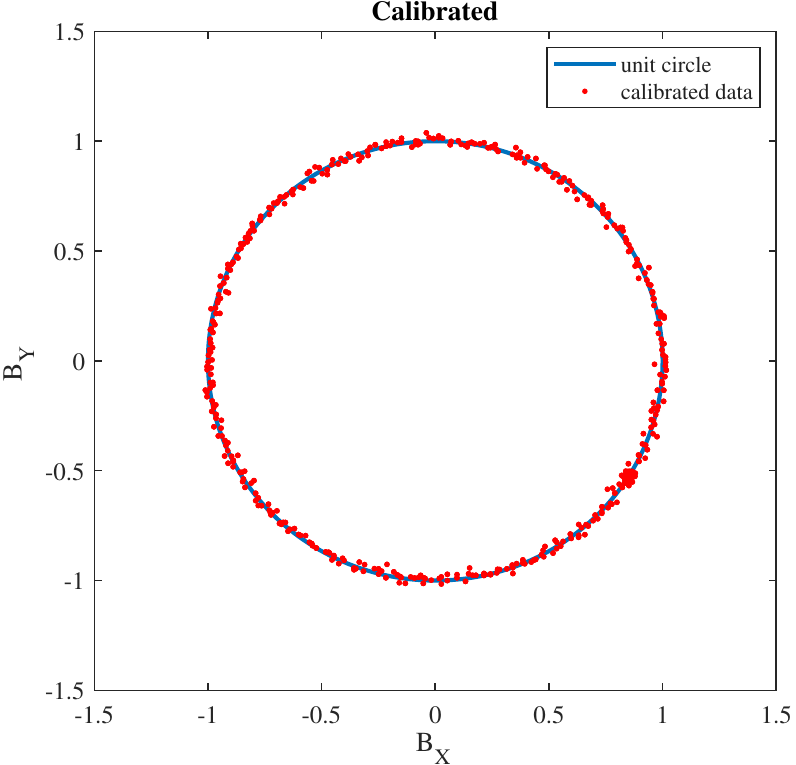}\label{fig:exp_cal_b}}
\caption{Measured magnetic-field data before and after calibration.}
\label{fig:exp_cal}
\end{figure}

The measured $x$- and $y$-direction magnetic-field data form an ellipse. This ellipse can be interpreted as a two-dimensional cross-section of the three-dimensional magnetic-field ellipsoid. Since the calibration excitation is one-dimensional rather than a full three-dimensional rotation, the measured data form a two-dimensional ellipse instead of a three-dimensional ellipsoid. Before calibration, the center of the ellipse does not coincide with the origin, its two axis lengths are different, and its principal axes are not aligned with the measurement coordinate axes. After calibration, the ellipse is expected to be transformed into a unit circle centered at the origin. The measured uncalibrated and calibrated data are shown in Fig.~\ref{fig:exp_cal} and the obtained circle deviation is 0.021.

\subsection{Test 1}
In the first test, the elevation angle of the motion table is increased by a few degrees. The table is then rotated through $360^\circ$, and the heading is estimated from the magnetometer measurements. The estimated heading is compared with the motor angle. The zero motor angle was adjusted to be as closely aligned as possible with the $0^\circ$ yaw reference. Since the residual motor-angle error and measurement noise are small, the motor azimuth is used as the reference magnetic heading.

\begin{figure}[!t]
\centering
\includegraphics[width=\columnwidth]{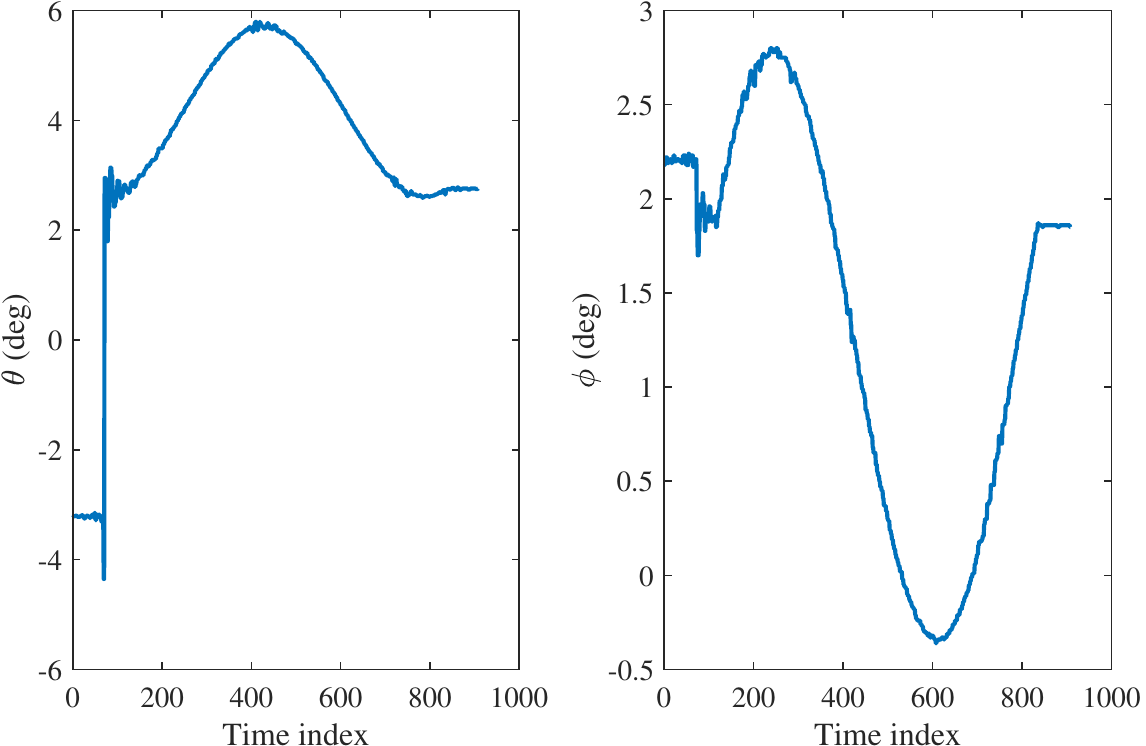}
\caption{Measured roll and pitch angles during experimental Test 1.}
\label{fig:test1_att}
\end{figure}

\begin{figure}[!t]
\centering
\includegraphics[width=\columnwidth]{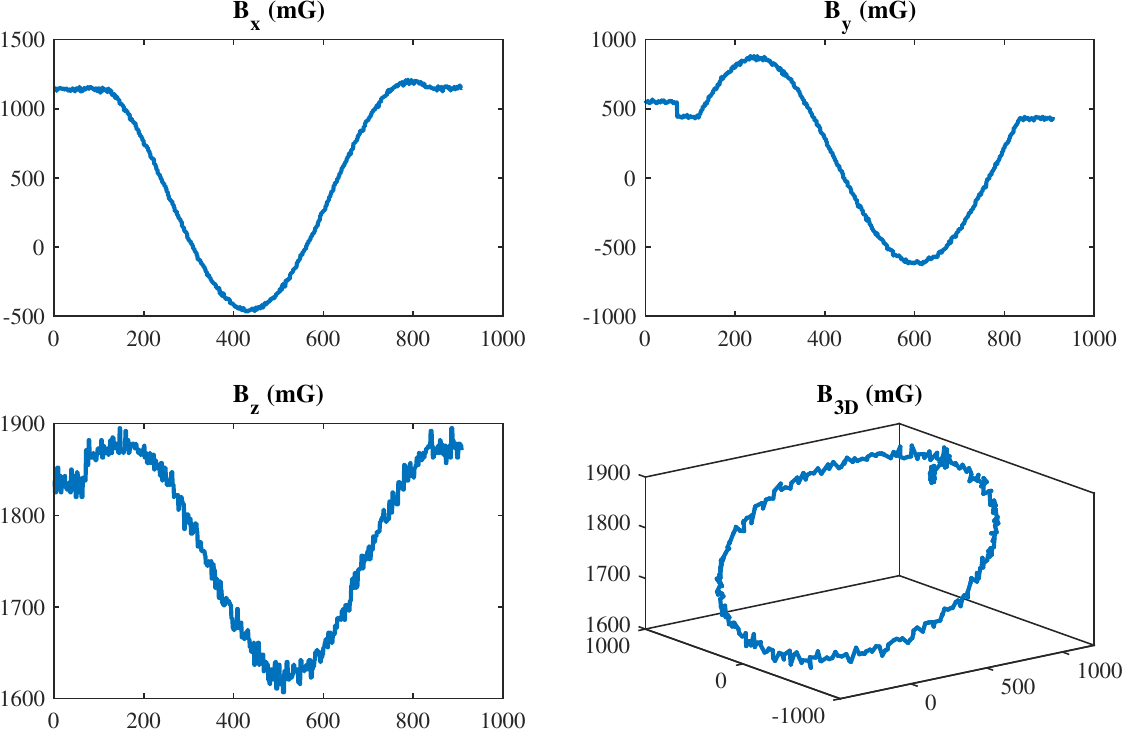}
\caption{Measured three-dimensional magnetic-field data during experimental Test 1.}
\label{fig:test1_mag}
\end{figure}

\begin{figure}[!t]
\centering
\includegraphics[width=\columnwidth]{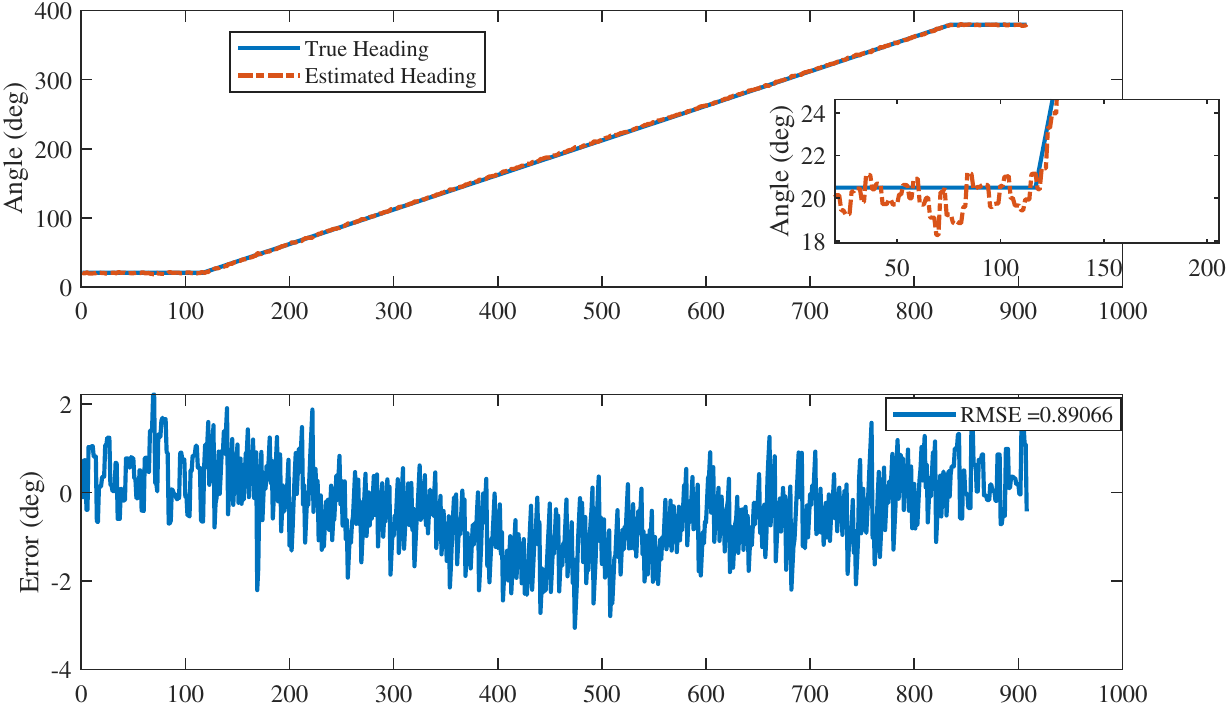}
\caption{Estimated and reference magnetic heading during experimental Test 1.}
\label{fig:test1_heading}
\end{figure}

The measured roll and pitch angles during this test are shown in Fig.~\ref{fig:test1_att}, and the corresponding three-dimensional magnetic-field measurements are shown in Fig.~\ref{fig:test1_mag}. The estimated magnetic heading is compared with the reference magnetic heading represented by the motor angle. The results are shown in Fig.~\ref{fig:test1_heading}. The proposed method accurately estimates the magnetic heading. The RMSE between the estimated heading and the motor angle is $0.89^\circ$.


\subsection{Test 2}
\begin{figure}[!t]
\centering
\includegraphics[width=\columnwidth]{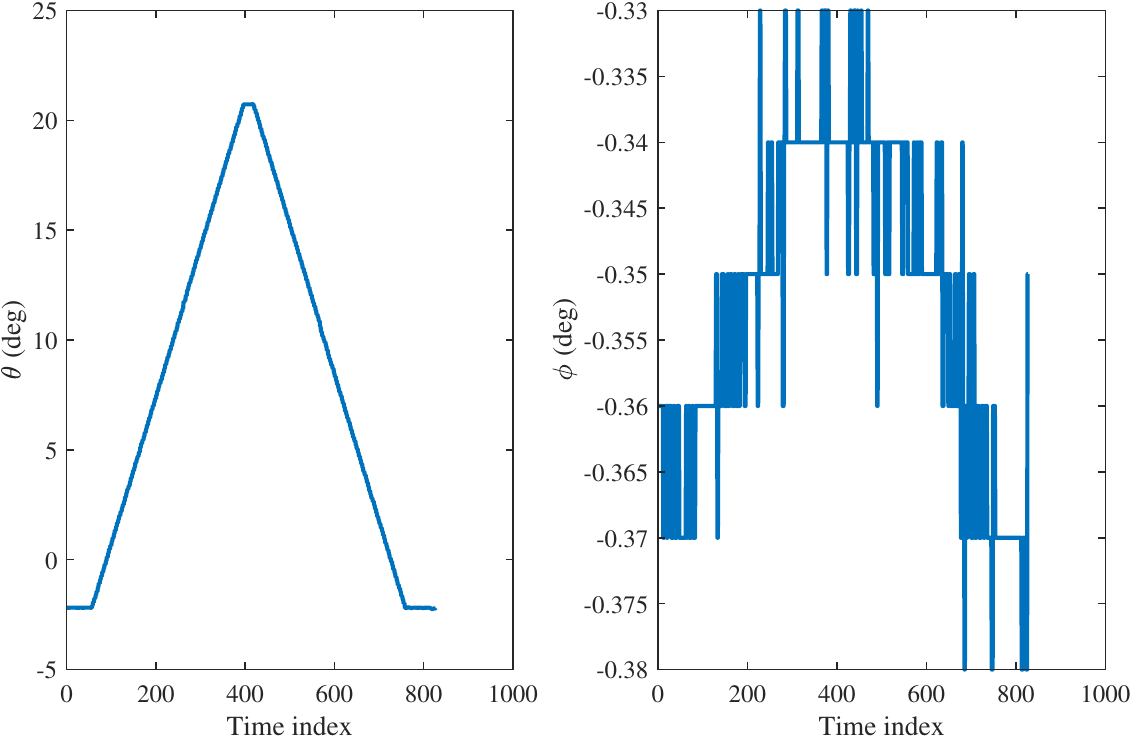}
\caption{Measured roll and pitch angles during experimental Test 2.}
\label{fig:test2_att}
\end{figure}

\begin{figure}[!t]
\centering
\includegraphics[width=\columnwidth]{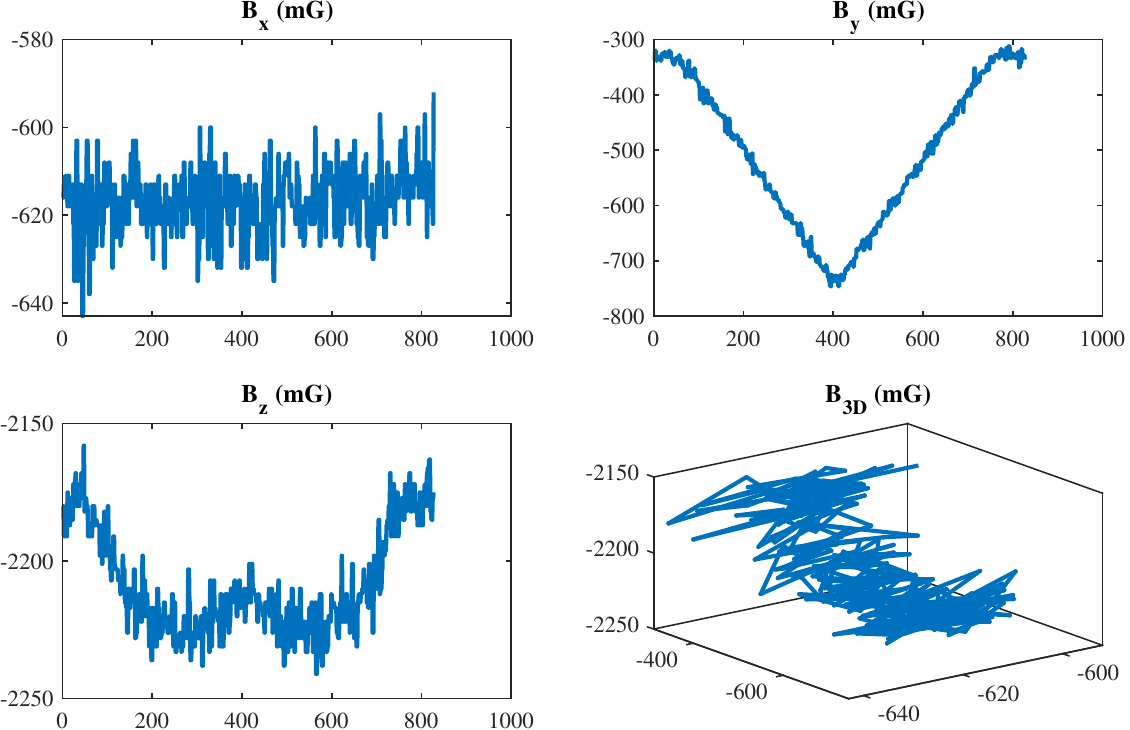}
\caption{Measured three-dimensional magnetic-field data during experimental Test 2.}
\label{fig:test2_mag}
\end{figure}

\begin{figure}[!t]
\centering
\includegraphics[width=\columnwidth]{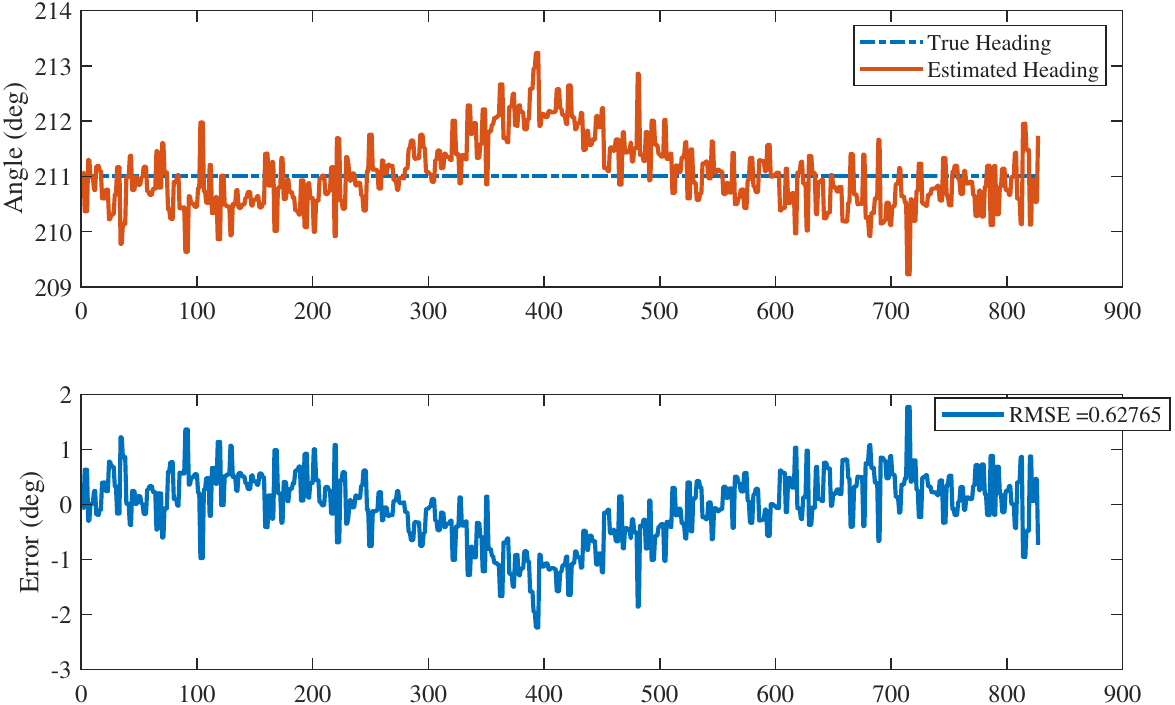}
\caption{Estimated and reference magnetic heading during experimental Test 2.}
\label{fig:test2_heading}
\end{figure}

In the second test, the motor is maintained at a fixed azimuth angle while its elevation angle is varied. This variation changes the pitch angle of the measurement module. If the module axis is not perfectly aligned with the motor axis, variations in roll angle can also occur. The corresponding elevation and roll variations and the three-dimensional magnetic-field measurements are shown in Figs.~\ref{fig:test2_att} and \ref{fig:test2_mag}, respectively. Finally, the estimated magnetic heading is compared with the azimuth angle of the motor. The comparison is shown in Fig.~\ref{fig:test2_heading}. The RMSE between the estimated heading and the motor angle is $0.63^\circ$. This result is particularly notable because the calibration procedure is performed under limited roll and pitch angles. Nevertheless, the obtained calibration parameters can be used not only for limited roll and pitch angles but also for pitch angles of up to approximately $22^\circ$, for which the proposed method still provides satisfactory heading estimation.


\section{Conclusion}
This paper presented a two-dimensional magnetometer calibration method for heading estimation using single-axis rotational excitation and measured roll and pitch information from inertial sensors. The proposed method addresses the practical problem of magnetometer calibration when a full three-dimensional calibration motion is not available or is undesirable for in-field and operational applications. In particular, the method uses two magnetic measurement channels and incorporates the available attitude information to compensate for the effects that arise from variations in roll and pitch during the calibration motion. The calibration parameters are obtained using a closed-form and non-iterative procedure, without requiring an iterative optimization algorithm or an initial parameter estimate.

The performance of the proposed method was evaluated through both numerical simulation and experimental measurements. In the simulation results, the method provided accurate heading estimation under the considered sensor-error conditions and remained effective when an asymmetric distortion matrix was used, despite the symmetric-matrix assumption adopted in the formulation. In the experimental evaluation, the RMS heading error was $0.89^\circ$ in the first test and $0.63^\circ$ in the second test. Moreover, although the calibration procedure was performed under limited roll and pitch angles, the calibrated sensor provided satisfactory heading estimation during the test stage for pitch angles of up to approximately $22^\circ$. These results demonstrate that the proposed two-dimensional calibration approach can provide accurate heading estimation while requiring only single-axis rotational excitation.

In this paper, the proposed method uses two-dimensional magnetometer calibration for heading estimation; however, the parameters of the complete distortion matrix are not fully extracted. In future work, the two-dimensional calibration framework can be extended by using magnetic-field measurements in all three directions to estimate the remaining parameters of the distortion matrix. In addition, future work can investigate the maximum heading accuracy achievable by a magnetometer for a given measurement-noise level. This can be addressed by deriving the Cram'er--Rao lower bound (CRLB) for heading estimation and comparing the performance of the methods with this theoretical bound.

\appendices
\section{Ellipse Fitting}
The ellipse-fitting procedure used in this work is based on least-squares fitting of a general conic to a set of two-dimensional measurement points. The same procedure is used in both calibration stages. The fit is accepted only when the resulting conic satisfies the ellipse condition $B^2-4AC<0$ and yields real, positive semi-axis lengths. For a set of $N$ measured points $(x_i,y_i)$, the conic equation is
\begin{equation}
Ax^2+Bxy+Cy^2+Dx+Ey+F=0.
\end{equation}
Define
\begin{equation}
\mathbf{p}=\begin{bmatrix}A&B&C&D&E\end{bmatrix}^{T},
\end{equation}
\begin{equation}
\mathbf{X}=\begin{bmatrix}
x_1^2&x_1y_1&y_1^2&x_1&y_1\\
x_2^2&x_2y_2&y_2^2&x_2&y_2\\\vdots&\vdots&\vdots&\vdots&\vdots\\x_N^2&x_Ny_N&y_N^2&x_N&y_N
\end{bmatrix}.
\end{equation}
The arbitrary conic scale is fixed by setting $F=-1$, so that $\mathbf X\mathbf p\simeq\mathbf 1$. This normalization is applicable to the ellipses encountered in the present calibration data. The least-squares estimate is
\begin{equation}
\widehat{\mathbf p}=\arg\min_{\mathbf p}\|\mathbf X\mathbf p-\mathbf 1\|_2^2.
\end{equation}
When $\mathbf X$ has full column rank, the solution is obtained by a standard least-squares solve; the normal-equation expression $\widehat{\mathbf p}=(\mathbf X^T\mathbf X)^{-1}\mathbf X^T\mathbf 1$ is shown only as an algebraic form and an explicit matrix inverse is not required in implementation. Here $\mathbf 1$ is an $N\times1$ vector of ones. The orientation angle of the fitted ellipse is
\begin{equation}
\phi=\frac{1}{2}\operatorname{atan2}(B,C-A).
\end{equation}
Let $c=\cos\phi$ and $s=\sin\phi$. The coefficients in the untilted coordinate system are
\begin{equation}
A'=Ac^2-Bcs+Cs^2,
\end{equation}
\begin{equation}
C'=As^2+Bcs+Cc^2,
\end{equation}
\begin{equation}
D'=Dc-Es,\qquad E'=Ds+Ec.
\end{equation}
The cross-product coefficient becomes zero in the rotated coordinate system. Completing the squares gives the center in the rotated coordinates,
\begin{equation}
X_0=-\frac{D'}{2A'},\qquad Y_0=-\frac{E'}{2C'},
\end{equation}
and
\begin{equation}
F''=-F+\frac{D'^2}{4A'}+\frac{E'^2}{4C'},\qquad F=-1.
\end{equation}
With the rotation convention used above, the ellipse center in the original measurement coordinates is
\begin{equation}
\begin{bmatrix}x_0\\y_0\end{bmatrix}
=
\begin{bmatrix}
c&s\\
-s&c
\end{bmatrix}
\begin{bmatrix}X_0\\Y_0\end{bmatrix}.
\end{equation}
For an accepted ellipse, $F''/A'>0$ and $F''/C'>0$, and the semi-axis lengths are
\begin{equation}
a=\sqrt{\frac{F''}{A'}},\qquad b=\sqrt{\frac{F''}{C'}}.
\end{equation}
If these positivity conditions are not satisfied, the conic fit is rejected as invalid for the calibration procedure. Thus, the five geometric parameters returned to the calibration algorithm are
\begin{equation}
a,\quad b,\quad \phi,\quad x_0,\quad y_0.
\end{equation}

\bibliographystyle{IEEEtran}
\bibliography{references}

\end{document}